\documentclass[preprint,aps,pra,superscriptaddress,showpacs,floatfix]{revtex4}
\usepackage{graphicx}
\usepackage{wrapfig}
\usepackage{dcolumn}
\usepackage{bm}
\usepackage{amsfonts}
\usepackage{amsmath}
\usepackage{cancel}
\usepackage{multirow}
\begin{document}
\title{The harmonic hyperspherical basis for identical particles 
without permutational symmetry}

\author{M. Gattobigio}
\affiliation{INLN, Universit\'e de Nice-Sophia Antipolis, CNRS, 1361 route des Lucioles, 06560 Valbonne, France }
\author{A. Kievsky}
\author{M. Viviani}
\affiliation{Istituto Nazionale di Fisica Nucleare, Largo Pontecorvo 3, 56100 Pisa, Italy}
\author{P. Barletta}
\affiliation{Department of Physics and Astronomy, University College London, 
Gower Street, London WC1 E6BT, United Kingdom}

\begin{abstract}
The hyperspherical harmonic basis is used to describe bound states in an
$A$--body system. 
The approach presented here is based on the
representation of the potential energy in terms of hyperspherical 
harmonic functions.
Using this representation, the matrix elements between the basis elements are
simple, and the potential energy is presented in a compact form, well suited
for numerical implementation.
The basis is neither  symmetrized nor antisymmetrized, as required in the case
of identical particles; however, after the diagonalization of the Hamiltonian
matrix, the eigenvectors reflect the symmetries present in it, and  the
identification of the physical states is possible, as it will be shown in
specific cases. 
We have in mind applications to atomic, molecular, and nuclear few-body systems
in which symmetry breaking terms are present in the Hamiltonian; their
inclusion is straightforward in the present method.  
As an example we solve the case of three and four particles interacting through
a short-range central interaction and Coulomb potential. 
\end{abstract}

\maketitle

\section{Introduction}

The Harmonic Hyperspherical (HH) method provides a systematic way to construct
an expansion basis for a system of $A$ particles. The $N$ Jacobi vectors
($N=A-1$) are transformed to the hyperradius $\rho$  plus $3N-1$ hyperangular
coordinates which are used to define the HH functions.  These functions are the
eigenfunctions of the hyperangular part of the Laplacian operator for a given
number of particles (see Ref.~\cite{fabre83} and references therein).

Applications of the HH method to describe bound states of $A=3,4$ nuclei
are well documented in the literature (for a recent review see Ref.~\cite{kiev08}).
In these applications the HH basis elements, extended to spin and isospin degrees
of freedom, have been combined in order to construct antisymmetric basis functions. 
In fact, the HH functions, as normally defined, do not have well defined properties
under particle permutation; this results from the selection of a particular
ordering of the particles in the definition of the Jacobi coordinates
and, as a consequence, of the  hyperangular
coordinates used to define the HH functions.
Changing the ordering of the particles, it is possible to define a new set of
Jacobi coordinates and, accordingly, HH functions depending on the hyperangular
variables obtained from this new set. To be noticed that the HH functions
defined using a particular choice of the Jacobi coordinates form a complete
basis.

The HH functions defined in one set
of Jacobi coordinates can be transformed to HH functions defined in another
set. In this transformation (permutation) the grand angular quantum number $K$,
which identifies a subset of HH functions, is conserved. For finite 
values of $K$, the dimension $N_K$ of this subset
is finite, and therefore a finite number of HH functions, having all the 
same value of $K$, are necessary to describe a HH function having 
the same value of $K$ but defined in a different Jacobi set.
The coefficients of the transformation can be collected in
a matrix having the dimension $N_K$ for each number
of particles. For $A=3$ these matrix elements  are the Raynal-Revai
coefficients~\cite{rr}. For $A>3$ the coefficients cannot be given in
a close form, and a few methods have been devised for their 
calculations~\cite{krivec90,viv98,novo94,efros95}. The knowledge of these 
coefficients allows for the construction of basis elements with well 
defined permutational symmetry. In fact, each subset defined by $K$
is invariant under particle permutation, as a consequence,
the constructions of basis elements with
that property is performed as linear combinations of HH
functions having the same value of $K$. Different schemes to construct
hyperspherical functions with an arbitrary permutational symmetry 
are given in Refs.~\cite{novo94,barnea95,novo97,novo98,barnea99}. Recently, a procedure
for constructing HH functions in terms of a single particle basis has 
been proposed in Ref.~\cite{timofeyuk:08}.

In problems in which the $A$-body system is composed by identical particles
the wave function of the system has to be completely symmetric or
antisymmetric in the case of bosons or fermions, respectively.
Considering a Hilbert space extended to spin and isospin degrees of freedom,
the construction of HH functions having well defined permutational 
properties allows for a reduction of the large degeneracy of the basis. 
In general the completely symmetric or antisymmetric
basis functions are a small part of the total Hilbert space. 
However, the difficulties of constructing HH functions with well defined
permutational symmetries increase with $A$ and $K$; therefore, the
preliminary step of constructing basis functions with well defined permutational
symmetry for $A$ particles could be sometimes very difficult 
to carry out.

In the present paper we investigate a different strategy. We intend to perform
the description of a $A$-body system using the HH basis defined on one set of
Jacobi coordinates, the reference set, and not having a well defined behaviour
under particle permutation. We will loose the advantage of using a reduced part
of the total Hilbert space; however, we will gain in simplicity in the
calculation of the matrix elements.  By including all HH basis elements up to a certain
grand angular momentum $K$, the diagonalization of the Hamiltonian matrix will
produce eigenvectors reflecting its symmetries. If the Hamiltonian commutes
with the group of permutations of $A$ objects, $S_A$, in the case of
non degenerated eigenvalues, the eigenvectors will
have a well defined permutation symmetry, and can be organized in accordance
with the irreducible representations
of $S_A$. Therefore, identifying those eigenvectors with the desired
symmetry, the corresponding energies can be considered variational estimates.
In particular, it will be possible to identify a
subset of eigenvectors and eigenvalues corresponding exactly to those that
would be obtained performing the preliminary symmetrization of the states. 
The disadvantage of this method results in the large dimension of the matrices to
be diagonalized. However, at present, different techniques are available to
treat (at least partially) this problem.

For a system interacting through a two-body potential $V(i,j)$, the potential
energy operator results in a sum over pairs. Its matrix elements
can be reduced to one term, let us say $V(1,2)$, times the number of pairs when 
symmetric or antisymmetric state functions are considered.
When HH functions without well defined permutation behaviour are used, the 
calculation of the potential energy operator cannot be reduced
to the computation of one term. So we have to face the problem of
computing the matrix elements of a general term $V(i,j)$ between
HH functions defined in the reference set of Jacobi coordinates in which the
distance $r_{ij}$ between particles $(i,j)$ has not a simple form. 

The calculation of  $V(i,j)$ in the reference set of Jacobi coordinates is
performed in two steps: (i) first, we use a property of the HH basis which 
allows to expand a general function of the coordinates  $(i,j)$ in terms of 
a subset of the basis called the potential basis (PB)~\cite{fabre83}; 
(ii) then, as for the case of a generic HH basis element, each PB element
is transformed to the HH basis defined in the reference set of Jacobi coordinates.
In the case of the PB, the transformation coefficients are known
analytically for each value of $K$ and for a general number of particles $A$.
In this way, each term $V(i,j)$ of the potential energy can be first 
expanded in the PB, and then transformed to HH functions defined in
the reference set. So, after this procedure, the potential energy will
be expressed in terms of HH functions. As we will see, the computation of the
matrix elements of the potential energy is now very simple since
it results in a combination of integrals of three HH functions.
A compact form suitable for a numeric treatment of the problem
will be given. 

The derivation and implementation of the final expression 
for the potential energy in the calculation of
bound states is the main subject of the present paper. As a simple
application, a system of three and four nucleons interacting through a
central potential will be analyzed. Different symmetries will appear
considering or not the Coulomb interaction between two protons. 
To be noticed that when antisymmetrized basis functions are used to
describe three or four nucleons, the presence of the Coulomb interaction
implies that states with total isospin $T=1/2,3/2$ (for $A$=3) and
$T=0,1,2$ (for $A=4$) have to be considered.
The extension of the Hilbert space to include these terms increases
the dimension of the problem resulting comparable
to that one in which the antisymmetrization of the basis is not performed.
Finally we would like to stress that the present paper is the first
step in a program devoted to applications of the HH basis to systems 
with $A>4$ interacting through realistic potentials.

The paper is organized as follows, section \ref{sec:hh} is devoted to
a brief description of the HH basis. In sections \ref{sec:pot} the expression
for the potential energy in terms of HH states are given. In section \ref{sec:application}
the results for the examples proposed are shown. Section \ref{sec:conclu}
includes a brief discussion of the results and the perspectives of the
present work.

\section{The Harmonic Hyperspherical basis}\label{sec:hh}
In this section we present a brief overview of the properties of the HH basis
following Ref.\cite{fabre83}.
We start with the following definition of the Jacobi
coordinates for an $A$ body system:

\begin{equation}
  \mathbf x_{N-j+1} = \sqrt{\frac{2 m_{j+1} M_j}{(m_{j+1}+M_j)m} } \,
                  (\mathbf r_{j+1} - \mathbf X_j)\,,
   \qquad
   j=1,\dots,N\,,
  \label{eq:jacobiCoordinates}
\end{equation}
where $m$ is a reference mass, $N=A-1$ and we have defined
\begin{equation}
  M_j = \sum_{i=1}^j m_i\,, \qquad \mathbf X_j = \frac{1}{M_j} \sum_{i=1}^j
  m_i\mathbf r_i \,.
  \label{}
\end{equation}
Let us note that if all the masses are equal, $m_i = m\,$,
Eq.~(\ref{eq:jacobiCoordinates}) simplifies to
\begin{equation}
  \mathbf x_{N-j+1} = \sqrt{\frac{2 j}{j+1} } \,
                  (\mathbf r_{j+1} - \mathbf X_j)\,,
   \qquad
   j=1,\dots,N\,.
  \label{eq:jc2}
\end{equation}
For a given set of Jacobi coordinates $\mathbf x_1, \dots, \mathbf x_N$, 
we can introduce the hyperradius $\rho$
\begin{equation}
  \rho = \bigg(\sum_{i=1}^N x_i^2\bigg)^{1/2}
   = \bigg(2\sum_{i=1}^A (\mathbf r_i - \mathbf X)^2\bigg)^{1/2}
   = \bigg(\frac{2}{A}\sum_{j>i}^A (\mathbf r_j - \mathbf r_i)^2\bigg)^{1/2} \,,
  \label{}
\end{equation}
and the hyperangular coordinates $\Omega_N$
\begin{equation}
  \Omega_N = (\hat x_1, \dots, \hat x_N, \phi_2, \dots, \phi_N) \,,
  \label{}
\end{equation}
with the hyperangles $\phi_i$ defined via
\begin{equation}
  \cos\phi_i = \frac{x_i}{\sqrt{x_1^2 + \dots + x_i^2}}\,,\qquad i=2,\dots, N\,.
  \label{}
\end{equation}
The radial components of the Jacobi coordinates can be expressed in terms of
the hyperspherical coordinates
\begin{equation}
  \begin{aligned}
    &x_N = \rho \cos\phi_N \\
    &x_{N-1} = \rho \sin\phi_N \cos\phi_{N-1} \\
    &\qquad\vdots \\
    &x_{i} = \rho \sin\phi_N \cdots \sin\phi_{i+1}\cos\phi_i\\
    &\qquad\vdots \\
    &x_{1} = \rho \sin\phi_N \cdots \sin\phi_{3}\sin\phi_2  \,. \\
  \end{aligned}
  \label{} 
\end{equation}
Using the above hyperspherical angles $\Omega_N$, the surface
element becomes
\begin{equation}
  d\Omega_N = \sin\theta_1\,d\theta_1\,d\varphi_1
               \prod_{j=2}^N \sin\theta_j\,d\theta_j\,d\varphi_j
               (\cos\phi_j)^2 (\sin\phi_j)^{3j-4} d\phi_j \,,
  \label{eq:surface}
\end{equation}
and the Laplacian operator 
\begin{equation}
  \Delta = \sum_{i=1}^N \nabla_{\mathbf x_i}^2 = \left(
  \frac{\partial^2}{\partial\rho^2} +
  \frac{3N-1}{\rho}\frac{\partial}{\partial\rho} +
  \frac{\Lambda_N^2(\Omega_N)}{\rho^2}\right) \,,
  \label{}
\end{equation}
where the $\Lambda_N^2(\Omega_N)$ is the generalization of the angular momentum
and is called grand angular operator.

The HH functions ${\mathcal Y}_{[K]}(\Omega_N)$ are the eigenvectors 
of the grand angular momentum operator

\begin{equation}
  \bigg(\Lambda_N^2(\Omega_N) + K(K+3N-2)\bigg) {\mathcal Y}_{[K]}(\Omega_N) =
  0 \,.
  \label{}
\end{equation}
They can be expressed
in terms of the usual harmonic functions $Y_{lm}(\hat x)$ and of the
Jacobi polynomials $P_n^{a,b}(z)$. In fact, the explicit expression for the HH
functions is
\begin{equation}
  {\mathcal Y}_{[K]}(\Omega_N) = 
    \left[\prod_{j=1}^N Y_{l_jm_j}(\hat x_j) \right] \left[ \prod_{j=2}^N
    \,^{(j)}\!{\mathcal P}_{K_j}^{l_j,K_{j-1}}(\phi_j)\right] \,,
  \label{eq:hh}
\end{equation}
where $[K]$ stands for the set of quantum numbers $l_1,\dots,l_N,m_1, \dots,m_N,
n_2, \dots, n_N$, the hyperspherical polynomial is
\begin{equation}
^{(j)}{\mathcal P}_{K_j}^{l_j,K_{j-1}}(\phi_j) = {\mathcal
N}_{n_j}^{l_j,K_j} (\cos\phi_j)^{l_j} (\sin\phi_j)^{K_{j-1}} P^{\nu_{j-1}, l_j +
1/2}_{n_j}(\cos2\phi_j) \,,
\end{equation}
where the $K_j$ quantum numbers are defined as
\begin{equation}
  K_j = \sum_{i=1}^j (l_i + 2n_i)\,, \qquad n_1 = 0\,, \qquad K \equiv K_N\,,
  \label{}
\end{equation}
and the normalization factor
\begin{equation}
  {\mathcal N}_{n_j}^{l_j,K_j} = \sqrt{\frac{2\nu_j \Gamma(\nu_j-n_j)\,n_j!}
  {\Gamma(\nu_j-n_j -l_j -1/2) \,\Gamma(n_j+l_j+3/2)}} \;\;\; ,
  \label{}
\end{equation}
with $\nu_j = K_j +3j/2-1$.  
The quantum number $K$ is also known as grand angular momentum.

The HH functions are normalized
\begin{equation}
  \int d\Omega_N \bigg({\mathcal Y}_{[K']}(\Omega_N)\bigg)^* {\mathcal Y}_{[K]}(\Omega_N)
   = \delta_{[K],[K']} \,,
  \label{eq:normalization}
\end{equation}
moreover, the HH basis is complete
\begin{equation}
  \sum_{[K]} \bigg({\mathcal Y}_{[K]}(\Omega_N)\bigg)^* {\mathcal
  Y}_{[K]}(\Omega'_N)  = \delta^{3N-1}(\Omega'_N - \Omega_N)\,.
  \label{}
\end{equation}

With the above definitions, the HH functions do not have well defined total
orbital angular momentum $L$ and $z$ projection $M$. It is possible to
construct HH functions having well defined values of $LM$ by coupling the
functions $Y_{l_jm_j}(\hat x_j)$. This can be achieved using different coupling
schemes. Accordingly we can define the following HH function
\begin{equation}
    {\mathcal Y}^{LM}_{[K]}(\Omega_N) =
    \bigg[Y_{l_1}(\hat x_1)\otimes\ldots\otimes Y_{l_N}(\hat x_N) \bigg]_{LM} 
    \left[ \prod_{j=2}^N
    \,^{(j)}{\mathcal P}_{K_j}^{l_j,K_{j-1}}(\phi_j)\right] \,,
\end{equation}
having well defined values of $LM$, although the particular coupling scheme 
is not indicated. The set of quantum numbers $[K]$ now includes the values of
$LM$ plus $N-2$ intermediate $l$-values instead of the $N$ magnetic
numbers $m_j$. When necessary, the explicit coupling scheme of the above
HH function will be given.

\subsection{Potential Basis}
If we have a function which depends only on the difference of two particle
positions, $f(\mathbf r_1 - \mathbf r_2)$, we can use a subset of the HH's to
expand that function, called the potential basis (PB) \cite{fabre83}.
Let's introduce the Jacobi coordinates such that $\mathbf x_N = \mathbf r_1 -
\mathbf r_2$; then the PB subset is defined by the following condition 
($\Omega_{12}\equiv(\hat x_N,\phi_N$) and $\Omega_N=(\Omega_{N-1},\hat x_N,\phi_N$))
\begin{equation}
  \Lambda_{N-1}^2(\Omega_{N-1}) {\cal P}^{l,m}_{2n+l}(\Omega_{12}) = 0\,,
  \label{}
\end{equation}
where $(n,l,m) \equiv (n_N,l_N,m_N)$, and by
\begin{equation}
  \Lambda_{N}^2(\Omega_{N}) {\cal P}^{l,m}_{2n+l} (\Omega_{12})= -K(K+3N-2)
  {\cal P}^{l,m}_{2n+l}(\Omega_{12})\,,
  \label{}
\end{equation}
with $K = l + 2n$. Thus, the PB is a subset of the HH's which depends only
on $(\hat x_N, \phi_N )$ variables, and which is specified by only three 
quantum numbers $n,l,m$, instead of the $3N-1$. The PB basis element has well defined
angular momentum $l$ and projection $m$.
The expression of the PB elements is:
\begin{equation}
  {\cal P}^{l,m}_{2n+l} (\Omega_{12}) = Y_{lm}(\hat x_N)
  (\cos\phi_N)^l P_n^{3(N-1)/2-1,l+1/2} (\cos 2\phi_N) Y_{[0]}(D-3)\,,
\end{equation}
where ($D=3N$)
\begin{equation}
  Y_{[0]}(D-3)=\left[\frac{\Gamma[(D-3)/2]}{ 2\pi^{(D-3)/2}}\right]^{1/2}
\end{equation}
is the normalization verifying
\begin{equation}
\int[Y_{[0]}(D-3)]^2 d\Omega_{N-1} = 1 \,.
\end{equation}
The surface element is conveniently written as
\begin{equation}
d\Omega_N=d\Omega_{N-1}d\Omega_{12}=d\Omega_{N-1}d{\hat x}_N
 d\phi_N (\cos\phi_N)^2(\sin\phi_N)^{3N-4} \,.
\end{equation}

We can extend the definition of the PB elements to depend on the 
coordinates of a general pair $(i,j)$ as 
${\cal P}^{l,m}_{2n+l} (\Omega_{ij})$. The coordinates
$\Omega_{ij}\equiv (\hat x_N,\phi_N)$
are now defined by a different ordering
of the particles entering in the Jacobi coordinates such that
$\mathbf x_N = \mathbf r_i - \mathbf r_j$. One important property
of the PB elements is the following. When a PB element is defined
in the space spanned by the coordinates $\Omega_{ij}$, 
its expression in terms of HH functions
defined in the reference set $\Omega_N$ 
corresponding to the ordering of the particles 
$1,2,....,N$ is known and results

\begin{equation}
  {\cal P}^{l,m}_{2n+l}(\Omega_{ij}) = \sum_{[K' = 2n+l]}
  \,^{(N)}C_{[K']}^{n,l}(\varphi^{ij})\, {\mathcal Y}^{lm}_{[K']}(\Omega_N)\,,
  \label{eq:rotateSide}
\end{equation}
where the coefficients $\,^{(N)}C_{[K']}^{n,l}(\varphi^{ij})$ are given
by the following relation
\begin{equation}
  \,^{(N)}C_{[K']}^{n,l}(\varphi^{ij}) = \left( \,^{(N)}P_{2n+l}^{l,0}(0)
  \sqrt{\frac{\Gamma(3(N-1)/2)}{2\pi^{3(N-1)/2}}}\, \right)^{-1}\,
  \int d\hat x\, Y^*_{lm}(\hat x)
  \,{\mathcal Y}^{lm}_{[K']}(\Omega^{ij}_{z}) \, .
  \label{eq:coefRotSide}
\end{equation}
The angles $\varphi^{ij}= \{\varphi^{ij}_N,\dots,\varphi^{ij}_2\}$ 
defined from the following kinematic rotation vector
\begin{equation}
  \mathbf z(\varphi^{ij}) = \mathbf x_N \cos\varphi^{ij}_N \mathbf + \mathbf
  x_{N-1}\sin\varphi^{ij}_N \cos\varphi^{ij}_{N-1} + \cdots +
  \mathbf x_1\sin\varphi^{ij}_N\sin\varphi^{ij}_{N-1}
\cdots\sin\varphi^{ij}_2 
  \label{eq:kvz}
\end{equation}
are chosen to verify $\mathbf z(\varphi^{ij}) =  \mathbf r_j - \mathbf r_i$.
The hyperangles $\Omega^{ij}_z$ are defined as
$\Omega^{ij}_z \equiv 
\{\hat x,\dots\hat x, \varphi^{ij}_N,\dots,\varphi^{ij}_2\}$,
with $\hat x$ repeated $N$-times. The particular form of the HH function
${\mathcal Y}^{lm}_{[K']}(\Omega^{ij}_{z})$ produces the
coefficients of Eq.(\ref{eq:coefRotSide}) to be independent of $m$. In
Eq.(\ref{eq:rotateSide}) the sum over all quantum numbers $[K']$ is 
limited by the condition $2n+l=K'$, showing that a PB basis element depending
on $\Omega_{ij}$ can be given as a linear combination of HH basis elements
having the same value of grand angular quantum number but depending on 
$\Omega_N$.  
A generic function $f(\mathbf r_i - \mathbf r_j)$ can be expanded in terms
of the PB as
\begin{equation}
 f(\mathbf r_i - \mathbf r_j)=\sum_{nlm} f_{nlm}(\rho) 
{\cal P}^{l,m}_{2n+l}(\Omega_{ij}) \,,
\end{equation}
with
\begin{equation}
\begin{aligned}
 & f_{nlm}(\rho) =\int d\Omega_{ij} f(\mathbf r_i - \mathbf r_j)
 \int d\Omega_{N-1}
[{\cal P}^{l,m}_{2n+l}(\Omega_{ij})]^* \\
 & = {1\over Y_0(D-3)}\int d\Omega_{ij} Y^*_{lm}\;
  (\cos\phi_N)^l P_n^{3(N-1)/2-1,l+1/2} (\cos 2\phi_N) f(\mathbf r_i - \mathbf r_j) \,.
\end{aligned}
\end{equation}
The functions $f_{nlm}(\rho)$ are the hyperradial multipoles. Introducing the 
transformation of Eq.(\ref{eq:rotateSide}) in the above expressions it is possible
to write a general function $f(\mathbf r_i - \mathbf r_j)$ in terms of HH functions
given in the reference set. We will use this property for the potential energy of
an $A$-body system.

\section{The potential energy in terms of HH functions}\label{sec:pot}

A local two-body interaction can be put in the form

\begin{equation}
  V(i,j) =\sum_l \bigg[ A_l(i,j) \otimes Y_l({\hat r}_{ij})\bigg]_0 V_l(r_{ij}) \,,
  \label{eq:twobodyp}
\end{equation}
where we use the compact notation
\begin{equation}
  \bigg[ A_{l_1}(i,j) \otimes Y_{l_2}({\hat r}_{ij})\bigg]_{LM} =
  \sum_{m_1m_2} (l_1m_1 l_2m_2|LM) A_{l_1m_1}(i,j) Y_{l_2m_2}({\hat r}_{ij})\,.
\end{equation}
$A_{lm}(i,j)$ is an operator independent of the coordinates $\mathbf
r_{ij}$, and the coupling with the spherical harmonics to zero in 
Eq.(\ref{eq:twobodyp}) shows that the
potential is a scalar in total space. We can use the PB elements to expand
each $l$-term of the expansion
\begin{equation}
V(i,j) =\sum_{ln} \left[ A_l(i,j) \otimes {\cal P}^l_{2n+l}
(\Omega_{ij})\right]_0 V^l_n(\rho) \,,
\end{equation}
where the functions $V^l_n(\rho)$ are obtained from the following
integral in the hyperangular space
\begin{equation}
\begin{aligned}
 & V^l_n(\rho)=\int d\Omega_{ij} V_l(r_{ij}) Y_{lm}({\hat r}_{ij})
 \int d\Omega_{N-1}
[{\cal P}^{l,m}_{2n+l}(\Omega_{ij})]^*  \\
 & = {1\over Y_0(D-3)}\int d\phi_N (\cos\phi_N)^{2+l}(\sin\phi_N)^{3N-4}
  P_n^{3(N-1)/2-1,l+1/2} (\cos 2\phi_N) V_l(r_{ij}) \,.
\end{aligned}
\end{equation}

The complete potential energy is 
\begin{equation}
\sum_{i<j}V(i,j)=\sum_{i<j}\sum_{ln}\left[
A_{l}(i,j)\otimes {\cal P}^l_{2n+l}(\Omega_{ij})\right]_0 V^l_n(\rho) \,\, .
\end{equation}

It would be convenient to have the potential energy expressed in the
coordinates defined by $\Omega$ (in the following we drop the suffix $N$ for
the reference set). To this end we transform the PB elements obtaining
\begin{equation}
 \begin{aligned}
\sum_{i<j}V(i,j) 
&=\sum_{ln} V^l_n(\rho) \sum_{[K' = 2n+l]}
\sum_{i<j}\,^{(N)}C_{[K']}^{n,l}(\varphi^{ij})
\, \left[A_l(i,j)\otimes {\mathcal Y}^{l}_{[K']}(\Omega)\right]_0  \\
&=\sum_{ln} V^l_n(\rho) \, {\mathcal G}^l_n(\Omega) \,,
 \end{aligned}
\label{eq:multex}
\end{equation}
where we have defined
\begin{equation}
{\mathcal G}^l_n(\Omega)= \sum_{[K' = 2n+l]}
\bigg(\sum_{i<j}\,^{(N)}C_{[K']}^{n,l}(\varphi^{ij})\bigg)
\, \left[A_l(i,j)\otimes {\mathcal Y}^{l}_{[K']}(\Omega)\right]_0 
\,\,\, .
\label{eq:simm1}
\end{equation}

The final form of
Eq.(\ref{eq:multex}) gives a general expression for the
potential energy in terms of the HH basis elements. In the case of
central potentials $l=0$ and $A_{lm}=1$, and the above expressions reduce to
(omitting the indices $l=0,m=0$)
\begin{equation}
 \begin{aligned}
\sum_{i<j}V(i,j) &=\sum_{i<j} \sum_n \bigg( \sum_{[K' = 2n]}
 \,^{(N)}C_{[K']}^{n}(\varphi^{ij})\, {\mathcal Y}_{[K']}(\Omega)
\bigg) V_n(\rho) \\
&=\sum_{n} V_n(\rho) \sum_{[K' = 2n]}
\bigg(\sum_{i<j}\,^{(N)}C_{[K']}^{n}(\varphi^{ij})\bigg)
\, {\mathcal Y}_{[K']}(\Omega)  \\
&=\sum_n V_n(\rho) \, {\mathcal G}_n(\Omega) \,,
 \end{aligned}
\label{eq:multex0}
\end{equation}
with
\begin{equation}
 \begin{aligned}
{\mathcal G}_n(\Omega) & = \sum_{[K'=2n]}
\bigg(\sum_{i<j}\,^{(N)}C_{[K']}^{n}(\varphi^{ij})\bigg)
{\mathcal Y}_{[K']}(\Omega)  \\
 & = \sum_{i<j} {\cal P}_{2n}(\Omega_{ij})  \,\, .
 \end{aligned}
\label{eq:simm2}
\end{equation}

The matrix elements 
of the potential energy between two different
HH basis elements result
\begin{equation}
\langle {\cal Y}^{L_1M_1}_{[K_1]}|\sum_{i<j}V(i,j)|
{\cal Y}^{L_2M_2}_{[K_2]}\rangle_\Omega  =
 \sum_{nl} V^l_n(\rho) 
 \langle {\cal Y}^{L_1M_1}_{[K_1]}|{\mathcal G}^l_n(\Omega)
|{\cal Y}^{L_2M_2}_{[K_2]}\rangle_\Omega  
\,\, .
\label{eq:cpot}
\end{equation}
The above expression represents an integral in the hyperangular space, and
shows the tensor product form between the
hyperradius and the hyperangular coordinates which is typical
using the HH basis. The matrix
elements of the operators ${\mathcal G}^l_n(\Omega)$ are independent
of the potential
\begin{equation}
 \begin{aligned}
 &\langle {\cal Y}^{L_1M_1}_{[K_1]}|{\mathcal G}^l_n(\Omega)|
{\cal Y}^{L_2M_2}_{[K_2]}\rangle_\Omega= 
\sum_{[K' = 2n+l]} 
\sum_{i<j}\,^{(N)}C_{[K']}^{n,l}(\varphi^{ij}) \\
&\times  \, \sum_m {(-1)^{l-m}\over \sqrt{2l+1}} A_{l-m}(i,j) 
\int\! d\Omega \; [{\cal Y}^{L_1M_1}_{[K_1]}\!(\Omega)]^*\;
{\cal Y}^{lm}_{[K']}(\Omega)
 \;{\cal Y}^{L_2M_2}_{[K_2]}(\Omega)  \;\; .
 \end{aligned}
\end{equation}
Each ${\mathcal G}^l_n$ is a combination of HH
functions with grand orbital momentum $K'=2n+l$, therefore
its matrix elements follow a triangular relation. In fact, given
$K_1$ and $K_2$, the values of $n,l$ to be considered in the sum of
Eq.(\ref{eq:cpot}) are
limited by the relation $|K_1-K_2|\le 2n+l \le K_1+K_2 $. A
triangular relation is also verified by the orbital angular momenta:
 $|L_1-L_2|\le l \le L_1+L_2$. 
Furthermore, the matrix elements of ${\mathcal G}^l_n$ includes 
the computation of integrals of three HH basis elements:
\begin{equation}
\int d\Omega \; [{\cal Y}^{L_1M_1}_{[K_1]}\!(\Omega)]^*
\;{\cal Y}^{lm}_{[K']}(\Omega) \;{\cal Y}^{L_2M_2}_{[K_2]}(\Omega)  \;\; .
\end{equation}
These integrals factorize in products of one-dimensional integrals
consisting of either three hyperspherical polynomials or three
spherical harmonics that can be obtained analytically or very
efficiently using quadratures. 

As shown in Eq.(\ref{eq:simm1}),
each function ${\mathcal G}^l_n(\Omega)$ is symmetric in the particle
indices, therefore its
corresponding eigenvectors will have well defined symmetry under particle
permutations. For example,
when $A_{lm}(i,j)=1$, $K_1=K_2=K$, $l=0$, 
implying $(L_1,M_2)=(L_2,M_2)=(L,M)$, and $2n=2K$, 
the following elements
\begin{equation}
\langle {\cal Y}^{LM}_{[K]}|{\mathcal G}_{K}(\Omega)|
 {\cal Y}^{LM}_{[K]}\rangle_\Omega =
\langle{\cal Y}^{LM}_{[K]}| \sum_{i<j} {\cal P}_{2n}(\Omega_{ij}) |
 {\cal Y}^{LM}_{[K]}\rangle_\Omega \,,
\end{equation}
form a matrix by varying all the quantum numbers in $[K]$ with
fixed values of $K$ and $L,M$. The dimension of the matrix is
given by all HH functions with grand angular quantum number $K$ 
coupled to $L,M$. Its eigenvectors, which are 
combinations of this family of HH functions, 
will have well defined permutational symmetry. This reflects the fact 
that each $K$-subset is invariant under particle permutations.
Therefore, the diagonalization of the above matrix
is a way to construct basis elements with well defined
permutational symmetry using HH functions with fix values of $K$ and $L$.

\section{Application to systems with $A=3,4$}\label{sec:application}

In the description of bound states in an $A$-body system it is
common to use basis elements having the required symmetry, symmetric
states for bosons or antisymmetric states for fermions. 
In the present section we will analyze the use of
the HH basis without the initial
symmetrization or antisymmetrization of the basis. Although the
basis elements have not the required symmetry, the eigenvectors of
the Hamiltonian will have a well defined symmetry reflecting 
the symmetries appearing in the Hamiltonian. Therefore,
among all eigenvectors and eigenvalues, 
the physical states have to be identified.

By taking opportune linear combinations of the
HH basis elements, specific symmetries under particle permutation can be
constructed for fixed values of $K$. Therefore two calculations, one in which
all HH states up to a maximum value of $K$ are considered and the other in which
states with a particular symmetry up to the same value of $K$ are considered,
produce the same eigenvectors and eigenvalues. Of course, in the first
calculation will appear eigenvectors and eigenvalues belonging to other
symmetries not present in the second calculation.  The simplification of
avoiding the initial basis symmetrization is counterbalanced by the 
larger dimension of the Hamiltonian matrix. 

Limiting the discussion to central potentials, Eq.(\ref{eq:cpot}) is
well suited for a direct application of the HH basis. 
Let us introduced the following orthonormal basis element
\begin{equation}
  \langle\rho\,\Omega\,|\,m\,[K]\rangle =
  \bigg(\beta^{(\alpha+1)/2}\sqrt{\frac{m!}{(\alpha+m)!}}\,
  L^{(\alpha)}_m(\beta\rho)
  \,{\text e}^{-\beta\rho/2}\bigg)
  {\cal Y}^{LM}_{[K]}(\Omega)  \,,
  \label{mhbasis}
\end{equation}
where $L^{(\alpha)}_m(\beta\rho)$ is a Laguerre polynomial with
$\alpha=3N-1$ and $\beta$ a variational non-linear parameter. 
We will discuss the case $L=0$ for $A=3,4$.
The HH basis elements are
\begin{equation}
 {\cal Y}_{[K]}(\Omega)= {}^{(2)}\!P_K^{l,K_1}(\phi)  \;
  \bigg[Y_l({\hat x}_1)\otimes Y_l({\hat x}_2)\bigg]_0 
\end{equation}
for $A=3$, and
\begin{equation}
   {\cal Y}_{[K]}(\Omega)= {}^{(2)}\!P_{K_2}^{l_2,K_1}(\phi_2)\;
  {}^{(3)}\!P_{K}^{l_3,K_2}(\phi_3) \;
  \bigg[[Y_{l_1}({\hat x}_1)\otimes Y_{l_2}({\hat x}_2)]_{l_3}
\otimes Y_{l_3}({\hat x}_3)\bigg]_0 
\end{equation}
for $A=4$.
The corresponding matrix elements of the
Hamiltonian are 
\begin{equation}
 \langle \,m'\,[K']|H|\,m\,[K]\rangle =
 \langle \,m'\,[K']|T+V|\,m\,[K]\rangle  \,\, .
\end{equation}
The matrix elements of the potential energy corresponding to
the $\Omega$-space have been discussed in the previous section.
Integrating also on $\rho$-space they result
\begin{equation}
 \langle \,m'\,[K']|V|\,m\,[K]\rangle  = \sum_n
 \langle \,m'|V_n(\rho)|\,m \rangle_\rho  
 \langle \,[K']|{\mathcal G}_n(\Omega)|\,[K]\rangle_\Omega
 \equiv
 \sum_n (V^n_{m'm})( {\mathcal G}^n_{[K'][K]}) \,\, .
\end{equation}
The matrix elements of the kinetic energy are the following
\begin{equation}
  \begin{aligned}
    T_{K' m'; K m} &= \langle m'\,[K']\,|\,
    -\frac{\hbar^2}{m} \sum_{i=1}^{N}  \nabla_{\mathbf x_i}^2
   | m\,[K]\rangle  \\
   &= -\frac{\hbar^2}{m} \langle m'\,[K']\,|\,
  \frac{\partial^2}{\partial\rho^2} +
  \frac{N-1}{\rho}\frac{\partial}{\partial\rho} +
  \frac{\Lambda_N^2(\Omega)}{\rho^2}
   | m\,[K]\rangle \\
   &= -\frac{\hbar^2\beta^2}{m} \delta_{[K],[K']} T_{m',m}^K \\
    & =-\frac{\hbar^2\beta^2}{m} \delta_{[K],[K']} 
    [T^{(1)}_{m'm}-K(K+3N-2)T^{(2)}_{m'm}]\,,
 \end{aligned}
\end{equation}
with
\begin{equation}
  \begin{aligned}
 & T^{(1)}_{m'm} = \frac{1}{4}\delta_{m,m'} 
 + \sqrt{\frac{m'!}{(\alpha+m')!}}\,\sqrt{\frac{m!}{(\alpha+m)!}} \\
 & \times \int_0^\infty x^\alpha\,{\text e}^{-x} dx\, 
L^{(\alpha)}_{m'}(x) \bigg[ \bigg(
   - \frac{\alpha+2m}{2x} - \frac{m}{x^2} \bigg)L^{(\alpha)}_{m}(x)
   +   \frac{m+\alpha}{x^2} L^{(\alpha)}_{m-1}(x)(1-\delta_{m,0})\bigg]\,,
  \end{aligned}
    \label{}
\end{equation}
and
\begin{equation}
  T^{(2)}_{m'm} = 
\sqrt{\frac{m'!}{(\alpha+m')!}}\,\sqrt{\frac{m!}{(\alpha+m)!}}\,
    \int_0^\infty x^\alpha\,{\text e}^{-x} dx\, L^{(\alpha)}_{m'}(x)\bigg(
    \frac{1}{x^2} \bigg)L^{(\alpha)}_{m}(x)\,.
\end{equation}
Using the properties of the Laguerre polynomials, these integrals can be
calculated analytically.

Therefore, the matrix elements of the Hamiltonian are sums of tensor products
of two matrices, one calculated on $\rho$-space, depending on indices $m,m'$,
and one calculated on $\Omega$-space, depending on the indices
$[K],[K']$
\begin{equation}
  \langle \,m'\,[K']|H|\,m\,[K]\rangle = -\frac{\hbar^2\beta^2}{m}
 ( T^{(1)}_{m'm}-K(K+3N-2) T^{(2)}_{m'm}) \delta_{[K'][K]}
 + \sum_n (V^n_{m'm})( {\mathcal G}^n_{[K'][K]}) \,\, .
\label{eq:hmm}
\end{equation}
If we introduce the diagonal matrix $D$ such that
$\langle [K']\,|\,D\, | [K]\rangle = \delta_{[K],[K']} K(K+3N-2)$, and the identity
matrix $I$ in $K$-space, we can rewrite the Hamiltonian schematically as
\begin{equation}
  H = -\frac{\hbar^2\beta^2}{m} (I \otimes {}^{(1)}T + D \otimes {}^{(2)}T) 
  + \sum_n {\cal G}_{n}\otimes V_n\,,
  \label{eq:schemH}
\end{equation}
in which the tensor product character of the expression is explicitly
given. 
A scheme to diagonalize such a matrix  is given in the
Appendix~\ref{app:product}.

In the following we give results for nucleon systems with $A=3,4$ using the
Volkov potential
\begin{equation}
 V(r)=V_R \,{\rm e}^{-r^2/R^2_1} + V_A\, {\rm e}^{-r^2/R^2_2}
\end{equation}
with $V_R=144.86$ MeV, $R_1=0.82$ fm, $V_A=-83.34$ MeV, and $R_2=1.6$ fm. The
nucleons are considered to have the same mass chosen to be equal to the
reference mass $m$ and corresponding to
$\hbar^2/m = 41.47~\text{MeV\,fm}^{-2}$.
With this parametrization of the potential, the
two-nucleon system has a binding energy of $0.54592$ MeV.

This potential has been used several times in the literature making its
use very useful to compare different methods 
\cite{barnea99,varga95,viviani05,timo02}. The results will be obtained
after a direct diagonalization of the Hamiltonian matrix of
Eq.(\ref{eq:hmm}) including $m_{max}+1$ Laguerre polynomials with a fix
value of $\beta$, and all
HH states corresponding to maximum value of the grand angular momentum
$K_{max}$. The scale parameter $\beta$ can be used as a non-linear
parameter to study the convergence in the index $m=0,1,\ldots,m_{max}$, with
$m_{max}$ the maximum value considered. In
the present analysis the convergence will be studied with respect to
the index $K_{max}$, therefore, the number of Laguerre polynomials at
each step, $m_{max}+1$, will be sufficiently
large to guarantee independence from $\beta$ of the physical eigenvalues
and eigenvectors.

In Table~\ref{table:sym} we show the different
symmetries of the eigenvectors and the corresponding eigenvalues, for
$A=4$, in the particular case in which the Hamiltonian matrix
has been diagonalized for $m_{max}=0$, $\beta=2$ fm$^{-1}$,
and $K_{max}=6$. In this case the total dimension of the matrix is 56
with 32 ``even'' elements, corresponding to even values of $l_3$, and 24 ``odd'' 
elements corresponding to odd values of $l_3$. 
In particular there are 6 totally symmetric states, irreducible representation
[4] using the Yamagouchi symbol, 2 totally antisymmetric states, $[1^4]$,
8 states belonging to the three-dimensional irreducible representation $[3\,1]$, 
6 states belonging  to the two-dimensional irreducible representation $[2^2]$, and
4 states belonging to the three-dimensional irreducible representation $[2\,1^2]$.
The lowest eigenvalue of each irreducible representation is given in the table.
In the last two columns of the table, the eigenvalues are reported
considering separately even and odd basis elements.
Symmetric states are formed exclusively by even-basis element whereas
antisymmetric states are formed exclusively by odd-basis elements. 
The three
mixed symmetries, one two-dimensional and the other two three-dimensional,
show degenerate eigenvalues. In order to distinguish between the two three-dimensional
mixed symmetries, we observe that the three degenerate eigenvalues
divide differently in even and odd elements. The mixed
symmetry $[3\,1]$ is twice degenerate 
when the expansion basis is restricted to even states whereas the mixed
symmetry $[2\,1^2]$ is not. Therefore performing two different diagonalizations,
one using a restricted basis considering only even states and one considering
only odd states, all the symmetries can be identified. 

Furthermore, we can see from Table~\ref{table:sym} that a bound state appears 
in correspondence to a symmetric state. The fact that only one spatial
symmetry is present in the bound state is a direct consequence of using a 
central potential. The final antisymmetrization of the state, as 
required in the case of four nucleons, is performed
by multiplying the spatial symmetric wave function by the corresponding 
spin functions,
singlet spin states $S_{12}=0$ for the two protons labelled $(1,2)$ and 
$S_{34}=0$ for the two neutrons labelled $(3,4)$. In the
case of using the isospin formalism, the spatial symmetric state is multiplied
by a four nucleon antisymmetric spin-isospin function having total spin $S=0$ 
and total isospin $T=0$. 

In Tables~\ref{table:three} and \ref{table:four} the convergence of the ground 
state binding energies for $A=3,4$ are given as 
a function of $K_{max}$, respectively. 
In the last column the point Coulomb interaction between the two protons,
labelled as particles $(1,2)$, has been considered. 
In the case without the Coulomb potential, spatial component of the ground state
is completely symmetric. When the Coulomb potential is taken into account
this component is symmetric with respect to particles $(1,2)$.
For $A=4$, it is also symmetric with respect to the particles $(3,4)$, the two neutrons.
In this case it is convenient to introduce the $H$-type Jacobi coordinates
(for a recent application see Ref.~\cite{vijande}):
\begin{equation}
  \begin{aligned}
    \mathbf x_3 &= \mathbf r_2 - \mathbf r_1 \\
    \mathbf x_2 &= \frac{\mathbf r_4 + \mathbf r_3}{\sqrt{2}}
    - \frac{\mathbf r_2 + \mathbf r_1}{\sqrt{2}}\\
    \mathbf x_1 &= \mathbf r_4 - \mathbf r_3  \,\, ,
  \end{aligned}
  \label{}
\end{equation}
and construct HH basis elements based on this type of coordinates. These HH
functions are linear combinations of the HH function based on the $K$-type 
coordinates, introduced in Eq.(\ref{eq:jc2}) and used in the previous sections,
at fixed values of the grand angular quantum number $K$. As example,
in Table~\ref{table:four} the two different types of Jacobi coordinates have
been considered. The dimension of the bases indicated corresponds to taking into
account even basis elements which are the only ones entering in the
construction of the bound states. As stated before, for the $K$-type Jacobi
coordinates this means to take even values of $l_3$. For the $H$-type, both
$l_1$ and $l_3$ are taken even. The dimension of the problem for obtaining the
eigenvalue at $K_{max}=30$ results to be 72 for $A=3$, and 7872 (4056) for
$A=4$, using the $K$-type ($H$-type). The use of the $H$-type Jacobi
coordinates reduces the dimension of the problem by nearly a factor of two.

The calculations corresponding to the two different types of coordinates
differ in the set of angles $\varphi^{ij}$ defined in
Eq.(\ref{eq:kvz}) reflecting the different way of defining the
interparticle distances in both cases. 
In the case in which the symmetric states are identified and constructed
before diagonalization, the dimension is reduced to
27 for $A=3$ and around 600 for $A=4$. We observe a considerable reduction 
in the dimension of the eigenvalue problem for the symmetrized
basis. However the computational cost of constructing HH states
with specific permutational symmetry has to be compared to the 
simplicity in constructing the matrix elements of ${\cal G}_n$ and
in solving the system of Eq.(\ref{eq:schemH}).
To be noticed that the
results using the symmetrized HH basis of Ref.\cite{barnea95,viviani05}
coincide with the results presented here for each value of $K_{max}$.
For the sake of comparison, the results using the stochastic
variational method (SVM)\cite{varga95} are shown in the table. 

When the Coulomb potential between protons is included
the system can be treated as composed by two different species,
the protons and the neutrons, having different interactions and
slightly different masses. Using the complete HH basis
this cause no extra difficulties since the following term can be
added to the Hamiltonian
\begin{equation}
 \sum_n (V^{c,n}_{m'm})( {\mathcal F}^n_{[K'][K]}) \,\, ,
\end{equation}
where
$V^{c,n}_{m'm}$ are the hyperradial matrix elements of the
Coulomb potential multipoles and
${\mathcal F}^n_{[K'][K]}$ is a matrix equivalent to 
${\mathcal G}^n_{[K'][K]}$ with the only difference that the
sum over $(i,j)$ is limited to protons.
This term has the tensor product form and therefore the Hamiltonian reads:
\begin{equation}
  H = I \otimes {}^{(1)}T + D \otimes {}^{(2)}T 
  + \sum_n {\cal G}_{n}\otimes V_n +\sum_n{\cal F}_n\otimes V^c_n \,.
  \label{eq:schemH2}
\end{equation}
In the above equation protons and neutrons are assumed to interact 
with the same short range potential. For realistic potentials this is not the
case and this assumption can be relaxed dividing the potential 
energy in three parts, one for the interaction between protons,
one for the interaction between neutrons and one for the interaction
between protons and neutrons. To be noticed that using the complete
HH basis the dimension of the problem does not change by distinguishing
protons and neutrons or not. However, as is a common procedure,
it is possible to treat the
system as composed by identical particles using the isospin formalism.
The Coulomb potential breaks the isospin symmetry and the use of 
antisymmetric states requires
the inclusion of different isospin components in the wave function.
The $A=3$ bound state will have isospin $T=1/2,3/2$ components 
whereas the $A=4$ bound state will have $T=0,1,2$ components. 
After including all these components the two procedures, one using the 
complete HH basis and the other using antisymmetrized states, will 
produce the same eigenvalues.
An example for this case is given in Table IV in which the results
for $A=3,4$ using antisymmetric basis states,
including the different isospin components, are shown. For $A=3$, the 
$T=1/2$ component is by far the most important one; however, the exact result
is obtained after including both components, $T=1/2$ and $3/2$. To be
noticed that the dimension of the basis using antisymmetrized HH
states with isospin components $T=1/2,3/2$ is the same of the
complete HH states using even basis elements. Therefore in this case
the preliminary antisymmetrization of the basis is not convenient.
For $A=4$ the $T=0$ component is by far the most important; however,
the exact result is obtained after including the three isospin
components $T=0,1,2$. In this case the dimension of the basis using
even HH states up to $K_{max}$ is greater than that using antisymmetrized
basis elements, since in the symmetrization with respect the two
neutrons is not included automatically in the even HH states and has to
be constructed by the diagonalization procedure. However the difference
in the dimension of the two cases is considerably 
reduced with respect to the case
in which the Coulomb potential was not included. 

The equivalence of the last columns of Tables II and III with columns third and
sixth of Table IV illustrates the simplicity of treating symmetry breaking terms 
using the HH basis without permutational symmetry.

\section{Conclusions}\label{sec:conclu}

In this work we have presented a direct use of the HH basis in the description
of a $A$-body system. The basis has neither been symmetrized nor
antisymmetrized as required by a system of identical particles. However, the
eigenvectors of the Hamiltonian have well defined permutation symmetry. Among
all the eigenvectors, the physical ones can be identified. The benefit of the
direct use of the HH basis is based on a particular simple form used to
represent the potential energy. Each term of the two-body potential $V(i,j)$
has been expanded in the potential basis and then expressed in terms of the HH
basis, defined in the reference set, by using the corresponding transformation
coefficients. 
These coefficients are known
for each value of $K$ and for a general number of particles $A$.
Once the potential has been expressed in terms of the HH basis,
it results in a sum of tensor product terms originated from the
separation of the hyperradial and the hyperangular coordinates
inherent to the method. Moreover, the kinetic
energy can be put in a tensor product form too. Therefore, the
matrix representation of the
Hamiltonian is expressed as a sum of tensor product matrices, and this 
particular form can be diagonalized very efficiently using the
technique given in the Appendix. 
As a test case we have studied 
three and four nucleons interacting through a central potential,
the Volkov potential, used many times in the literature. We have
shown how the symmetries are present in the spectrum and can be
identified. The symmetric and antisymmetric states appear as
singlets, whereas the mixed symmetries appear as multiplets. We
have identified all symmetries by dividing the spectrum in
even and odd components. In the studied cases, only one bound
state appears for $A=3$ and $4$ corresponding to a symmetric
state. To be noticed that if the potential depends on the spin-
isospin degrees of freedom, the Hamiltonian will still present
the tensor product form in the hyperradial, hyperangular, spin
and isospin spaces. 

For $A=4$ we have solved the problem using two different types
of Jacobi coordinates, namely the $K$-type corresponding to a 3+1
configuration, and the $H$-type corresponding to a 2+2 configuration.
The calculations using one or the other set differ in the values
of the angles $\varphi^{ij}$, which can be considered as input
parameters. Therefore the method gives a systematic 
way of introducing the different types of Jacobi coordinates.
The convenience of selecting one specific type is related to its
capability to produce basis states having partially the required
symmetry with a reduction of the total dimension of the problem. 
In the cases presented here, the $A=4$ bound state is
constructed using basis elements based on the $K$-type Jacobi coordinates
with even values of $l_3$ or based on the $H$-type with both
$l_1$ and $l_3$ restricted to even values. The latter resulting in
a basis with a dimension smaller by a factor of two.

A further benefit of using the complete HH basis is obtained when
symmetry breaking terms are included in the Hamiltonian. The
complete basis will generate eigenvectors having specific
permutation symmetries reflecting the symmetries 
present in the Hamiltonian. The complexity of the numerical problem
does not increase when these terms are present. 
This is not the case when symmetrized or antisymmetrized
basis are used. For example, in the case of a nuclear system, the presence 
of charge symmetry breaking terms requires the extension of the basis 
to include all the isospin components. As a specific example,
here we have analyzed the case of the Coulomb interaction between
protons. The results using the HH basis without well defined permutation 
symmetry have
been compared to the case in which antisymmetrized HH basis have been used.
In the latter case the different isospin components entering in the
wave function have to be included, resulting in spatial components
having more than one symmetry. Accordingly the dimension of the
basis increases. To this respect,
the numerical effort to reduce the Hilbert space to subspaces with
specific permutation symmetry is discussed in Ref.\cite{barnea95}. 
When one spatial symmetry is required, as for example a completely
symmetric spatial state, the convenience of constructing symmetric
HH state is obvious. When several spatial symmetries are present in the wave
function, as in the case of an $A$-nucleus wave function,
the convenience of constructing HH states with different spatial symmetries
has to be compared to the capability of solving a large eigenvalue
problem, for example that one given in Eq.(\ref{eq:schemH}).

In Refs.~\cite{kiev93,kiev97,viviani05} the HH basis, used to describe
three- and four-nucleon bound states, is antisymmetrized in the
following way. The total wave function is expanded in 
angular-spin-isospin channels and, for each channel, it is written 
as a sum of Faddeev-like amplitudes, each of them antisymmetric in the 
pair $(i,j)$. In this way the total wave function results antisymmetric.
Then, each $(i,j)$-amplitudes is expanded in the HH basis defined
from Jacobi vectors corresponding to the different ordering of
the particles. As a consequence,
the amplitudes for the different channels are not orthogonal,
resulting in a non-orthogonal basis. For large values of $K$ the
non orthogonality of the basis could causes numerical instabilities.
In particular, for $A=4$, this problem is overcome performing
an orthonormalization of the basis using the Gram-Schmidt technique
with quadruple precision in the numerical treatment of the process. 
Therefore, the extension to $A>4$ systems appears to be difficult. On the other
hand, the direct use of the HH basis without antisymmetrization circumvents
this problem. 
Therefore,
the method presented here has to be considered a first step in
a program devoted to the application of the HH basis to systems with
$A>4$. Further works along this line are the extension of the method
to treat realistic interactions and the numerical implementation of 
the Hamiltonian of Eq.(\ref{eq:schemH}) to systems with $A=5,6$.
The extension of the method to treat three-nucleon interaction terms
is also possible. In fact, the transformation of the spatial part of a
three-nucleon interaction $W(i,j,k)$ in terms of HH functions constructed
in the reference set  can be performed using the algorithm developed
for the "triplet basis" in Ref.~\cite{viv98}.

\appendix
\section{Efficient matrix-vector product for tensor-product
matrices}\label{app:product}

The algorithm we used to diagonalize the Hamiltonian is an iterative one,
namely Lanczos's~\cite{ietl} . These kind of algorithms are useful whenever an efficient
matrix-vector product can be used, as in the case of sparse matrices; in the
specific calculation, we have the product between a tensor-product matrix $M = A_1
\otimes A_2$, and a vector $\mathbf v$
\begin{equation}
  \mathbf w = M\cdot \mathbf v = (A_1 \otimes A_2)\cdot \mathbf v,
  \label{}
\end{equation}
with $A_1$ a $n\times n$ matrix, $A_2$ a 
$m\times m$ matrix, and $\mathbf v$ a $(n\cdot m)$-dimensional vector.

The product is done in three steps: (i) first, the vector $\mathbf v$ is 
reshaped in a $m\times n$ matrix $V$; (ii) then, the following matrix products 
are performed 
\begin{equation}
  W = (A_1\cdot(A_2\cdot V)^T)^T\,;
  \label{}
\end{equation}
(iii) finally, the matrix $W$ is reshaped into the $(n\cdot m)$-dimensional vector
$\mathbf w$, which is the result of the multiplication. The above algorithm is easily
generalized to tensor-products of $k$-matrices~\cite{buis:96}.


\newpage

\begin{table}
  \caption{Lowest Volkov-energy eigenvalues of each irreducible representations
  of $S_4$ for the $N=4$ case, with $m_{max}=0$, $K_{max}=6$, and $\beta =
  2\,\text{fm}^{-1}$.  The multiplets are further identified as being symmetric
  or anti-symmetric under permutation of particles 1-2.}
  \label{table:sym}
  \begin{center}
    \begin{tabular}{|c|c|c|c|c|}
      \hline
      \multicolumn{2}{|c|}{Irreps} &Eigen's (MeV) & Sym(1-2) & AntiSym(1-2) \\
     \hline 
     [4] & 
     {\tiny
      \begin{tabular}{|c|c|c|c|}
        \hline
        1 & 2 & 3 & 4 \\
        \hline
      \end{tabular}
     } &
    -25.794 &-25.794 & \\
     \hline 
     \multirow{2}{*}{[$2^2$]}
     &\multirow{2}{*}{
     \tiny
    \begin{tabular}{|c|c|}
      \hline
      1 & 2  \\
      \hline 
      3 & 4 \\
      \hline
    \end{tabular}
     }
     & 27.680 & 27.680  & \\
     & & 27.680  & & 27.680  \\
     \hline
     \multirow{3}{*}{[3 1]}  &
     \multirow{3}{*}{
     \tiny
      \begin{tabular}{|c|c|c|}
        \hline
        1 & 2 & 3 \\
        \hline
        \multicolumn{1}{|c|}{4} & \multicolumn{2}{|c}{} \\
        \cline{1-1}
      \end{tabular}
     } 
     & 28.430 & 28.430 & \\
     & & 28.430 & 28.430 & \\
     & & 28.430 &  & 28.430  \\
     \hline
     \multirow{3}{*}{[$2 1^2$]}  &
     \multirow{3}{*}{
     \tiny
      \begin{tabular}{|c|c|}
        \hline
        1 & 2 \\
        \hline
        \multicolumn{1}{|c|}{3} & \multicolumn{1}{|c}{} \\
        \cline{1-1}
        \multicolumn{1}{|c|}{4} & \multicolumn{1}{|c}{} \\
        \cline{1-1}
      \end{tabular}
     } & 102.85 & 102.85 & \\
     & & 102.85 & & 102.85 \\
     & & 102.85 & & 102.85 \\
     \hline
     $\vdots$ & $\vdots$  & $\vdots$  & $\vdots$ & $\vdots$ \\
     \hline
     [$1^4$]  & 
     {\tiny
        \begin{tabular}{|c|}
          \multicolumn{1}{c}{}\\
          \hline
          1  \\
          \hline
          2  \\
          \hline
          3  \\
          \hline
          4  \\
          \hline
          \multicolumn{1}{c}{}\\
        \end{tabular}
     }  &199.56 & &199.56  \\
     \hline
     $\vdots$ & $\vdots$  & $\vdots$  & $\vdots$ & $\vdots$ \\
     \hline
    \end{tabular}
  \end{center}
\end{table}

\begin{table}[h!]
  \caption{Results for the Volkov's potential, as a function of $K_{max}$
  using 30
  Laguerre's polynomials, and $\beta=3$ fm$^{-1}$ for the three-body case.
  In the last column the results including the Coulomb potential are given.}
 \label{table:three}
  \label{table:volkov3p}
\begin{center}
  \begin{tabular*}{\linewidth}{@{\extracolsep{\fill}}c c c c}
  \hline
  \hline
  $K_{\text{max}}$ \rule{0pt}{12pt} &  $N_{\text{HH}}$   & 
  \multicolumn{2}{c}{$E$ (MeV)} \\
  \hline 
   0  & 1   & 7.7075 & 6.9926 \\
   10 & 12  & 8.4157 & 7.7083 \\
   20 & 36  & 8.4623 & 7.7566 \\
   30 & 72  & 8.4647 & 7.7693 \\
   40 & 121 & 8.4649 & 7.7694 \\
   \hline
   SVM\cite{varga95}  & 30  & 8.46   &      \\
   \hline
\end{tabular*}
\end{center}
\end{table}

\begin{table}[h!]
  \caption{$A=4$ results for the Volkov's potential, using 25
  Laguerre's polynomials, and $\beta=2\,\text{fm}^{-1}$. Two different
  types of Jacobi coordinates have been used.
  In the last two columns the results 
  without and with Coulomb potential, using independently 
  $K$- or $H$-type Jacobi coordinates, are given, respectively.
  At fixed value of $K_{\text{max}}$ the results, using either one
  or the other type of coordinates, coincide.
  }
  \label{table:four}
\begin{center}
  \begin{tabular*}{\linewidth}{@{\extracolsep{\fill}}c c c c c }
  \hline 
  \hline 
  $K_{\text{max}}$ \rule{0pt}{12pt} &  $N_{\text{HH}}$($K$-type)  &
  $N_{\text{HH}}$($H$-type)  & \multicolumn{2}{c}{$E$ (MeV)} \\
  \hline 
   0  & 1        & 1        & 28.580& 27.748\\
   10 & 136      & 78       & 30.278& 29.456 \\
   20 & 1547     & 819      & 30.416& 29.596 \\
   30 & 7872     & 4056     & 30.420& 29.599 \\
   \hline
   SVM\cite{varga95} & 50 & & 30.42 &        \\
   \hline
\end{tabular*}
\end{center}
\end{table}

\begin{table}[h!]
  \caption{Contributions to the bound state energies, for $A=3,4$,
   of the different isospin components using
 antisymmetrized HH functions}
  \label{table:isospin}
\begin{center}
  \begin{tabular*}{\linewidth}{@{\extracolsep{\fill}}c c c |c c c}
  \hline 
  \hline 
 & \multicolumn{2}{c}{$A=3$} &
 & \multicolumn{2}{c}{$A=4$} \\
  \hline 
 $K_{max}$ & $T=1/2$ & $T=1/2,3/2$  &$K_{max}$ & $T=0$ & $T=0,1,2$  \\
  \hline 
  0    &  6.9926   &  6.9926  &   0     & 27.748 & 27.748 \\
  10   &  7.7072   &  7.7083  &   10    & 29.453 & 29.456 \\ 
  20   &  7.7555   &  7.7566  &   20    & 29.594 & 29.596 \\
  30   &  7.7582   &  7.7593  &   30    & 29.596 & 29.599  \\
  40   &  7.7583   &  7.7594  &         &        &        \\
  \hline 
\end{tabular*}
\end{center}
\end{table}

\end{document}